\documentclass[letterpaper, 10 pt, conference]{ieeeconf}
\IEEEoverridecommandlockouts 
\usepackage{graphicx}
\usepackage{xcolor}
\usepackage{colortbl}

\title{\LARGE \bf
Haptic Feedback Relocation from the Fingertips to the Wrist for Two-Finger Manipulation in Virtual Reality
}

\author{Jasmin E. Palmer$^{1}$, Mine Sarac$^{2}$, Aaron A. Garza$^{1}$, and Allison M. Okamura$^{1}$ 
\thanks{This work was supported in part by a National Science Foundation Graduate Research Fellowship, a GEM Fellowship, a Stanford EDGE Fellowship, National Science Foundation grant 1830163, and Combat Capabilities Development Command-Soldier Center (CCDC-SC) grant W81XWH-20-C-0008 in collaboration with Triton Systems, Inc.}
\thanks{$^{1}$J. E. Palmer, A. A. Garza, and A. M. Okamura are with the Department of Mechanical Engineering, Stanford University, CA, 94305, USA {\tt\small \{jasminp, aarongs, aokamura\}@stanford.edu}}%
\thanks{$^{2}$M. Sarac is with the Department of Mechatronics Engineering, Kadir Has University, Istanbul, Turkey.  {\tt\small mine.sarac@khas.edu.tr}}%
}

\newcommand{\newtext}[1]{\textcolor{black}{#1}}
\begin{document}

\maketitle
\thispagestyle{empty}
\pagestyle{empty}

\begin{abstract}
Relocation of haptic feedback from the fingertips to the wrist has been considered as a way to enable haptic interaction with mixed reality virtual environments while leaving the fingers free for other tasks. We present a pair of wrist-worn tactile haptic devices and a virtual environment to study how various mappings between fingers and tactors affect task performance. The haptic feedback rendered to the wrist reflects the interaction forces occurring between a virtual object and virtual avatars controlled by \newtext{the} index finger and thumb. We performed a user study comparing four different finger-to-tactor haptic feedback mappings and one no-feedback condition as a control. We evaluated users' ability to perform a simple pick-and-place task via the metrics of task completion time, path length of the fingers and virtual cube, and magnitudes of normal and shear forces at the fingertips. We found that multiple mappings were effective, and there was a greater impact when visual cues were limited. We discuss the limitations of our approach and describe next steps toward multi-degree-of-freedom haptic rendering for wrist-worn devices to improve task performance in virtual environments.
\end{abstract}

\section{Introduction}

Many tactile haptic feedback devices that render forces from interactions with virtual objects \newtext{provide} forces directly to the glabrous skin of the fingertips, taking advantage of their mechanoreceptor density \cite{schorr2017fingertip, Leonardis2017, Chinello2018}. Because these devices are deliberately designed to cover the fingers and fingertips, they prevent users from simultaneously interacting with physical tools, which is detrimental for many mixed reality scenarios such as surgical training, guidance for assembly and repair, and remote collaboration. Leaving the fingers free of encumbrance can also facilitate optical predictive tracking to enhance the responsiveness of haptic feedback~\cite{SalvatoRAL2022}.

\newtext{Haptic feedback can be relocated from the fingertips to other locations on the body,  which would enable haptic interaction with mixed reality virtual environments while leaving the fingers free for other tasks or to simultaneously manipulate physical tools. We propose to provide feedback at the wrist or the forearm near the wrist.} Such relocation has been previously shown to improve task performance, user acceptance, and task enjoyment compared to scenarios with no haptic feedback~\cite{Pezent2019, Young2019, Moriyama2018, RaitorICRA2017}. Due to lower mechanoreceptor density on the wrist \cite{Johansson2009}, we aim to design wrist-worn haptic devices that provide meaningful feedback to the user -- but without the need \newtext{to create} perfectly realistic sensations. That is, the less accurate perception on the arm compared to the fingertips might make haptic illusions more feasible~\cite{CulbertsonAR2018}. 

We aim to understand the best practices of haptic relocation to the wrist during virtual manipulation tasks. We previously investigated the effects of feedback location on the wrist (dorsal vs.\ ventral vs.\ both sides)~\cite{sarac2019effects} and the rendered feedback direction (normal vs.\ shear) \cite{sarac2019haptic, Sarac2022}. Our results showed that participants’ perception was not affected by either feedback location or direction, as long as the participants observed noticeable, meaningful haptic feedback. In these studies, participants were asked to discriminate the stiffness property of virtual objects by pushing the objects toward the ground using the index finger. Thus, we only calculated the interaction forces acting on a single fingertip and rendered \newtext{haptic feedback} on the wrist. However, generic manipulation tasks should \newtext{be} performed by at least two fingers\newtext{, e.g.\ the} index finger and thumb\newtext{,} creating \newtext{opposing} forces acting on the objects. Even though one might assume that the interaction forces on both fingertips should be \newtext{equal at all times}, this would not correspond to scenarios in which the user \newtext{translates} or rotates a held object in the virtual environment due to inertial and gravitational effects. 

 In this paper, we created and tested various \newtext{finger-to-tactor} mappings that define the relationship between virtual environment interaction forces at the fingertips and rendered haptic feedback at particular wrist locations. The tactor is the \newtext{part of the} haptic device that presses into the skin, in our case near the wrist. We hypothesize that participants will perform a simple virtual manipulation task better when (1) fingertip forces are rendered on the wrist individually (rather than in combination or using a single fingertip force), and (2) index \newtext{finger} and thumb forces are mapped to the dorsal and ventral sides on the wrist, respectively (rather than to the ventral and dorsal sides, respectively). To test this, we calculated the virtual environment forces acting on the user’s index finger and thumb and used those forces to generate haptic feedback via a pair of wrist-worn tactile haptic devices mounted on the same wrist. We conducted a user study to investigate the impact of \newtext{finger-to-tactor} mapping on user performance during \newtext{a simple} virtual manipulation task. The main contribution of this work is an understanding of how the mappings of fingertip haptic interactions to locations on the wrist affect manipulation of virtual objects, to guide future design and rendering for wrist-worn haptic devices.

\section{Experimental Setup}

\subsection{Device Design}
\label{device_design_sec}

A pair of one-degree-of-freedom (1-DoF) wrist-worn \newtext{tactile} haptic devices provide skin deformation normal to the user’s wrist \newtext{via a tactor} to display simulated interaction forces. Each tactor is actuated via a \newtext{lightweight 6.2 g HXT500 micro servo motor by Hextronik}. \newtext{The} servo motor \newtext{is} attached to a rack and pinion mechanism \newtext{which was 3D printed}. \newtext{The rack has a maximum usable extension of 15 mm and the pinion has a pitch radius of 5.75 mm.} The servo motor is \newtext{open-loop} position controlled via digital I/O using a Sensoray 826 PCI Express board \newtext{at constant speed with a PWM frequency of 50~Hz}. 

\begin{figure}[t]
\vspace{1.5mm}
  \centering
  \includegraphics[width=0.8\columnwidth]{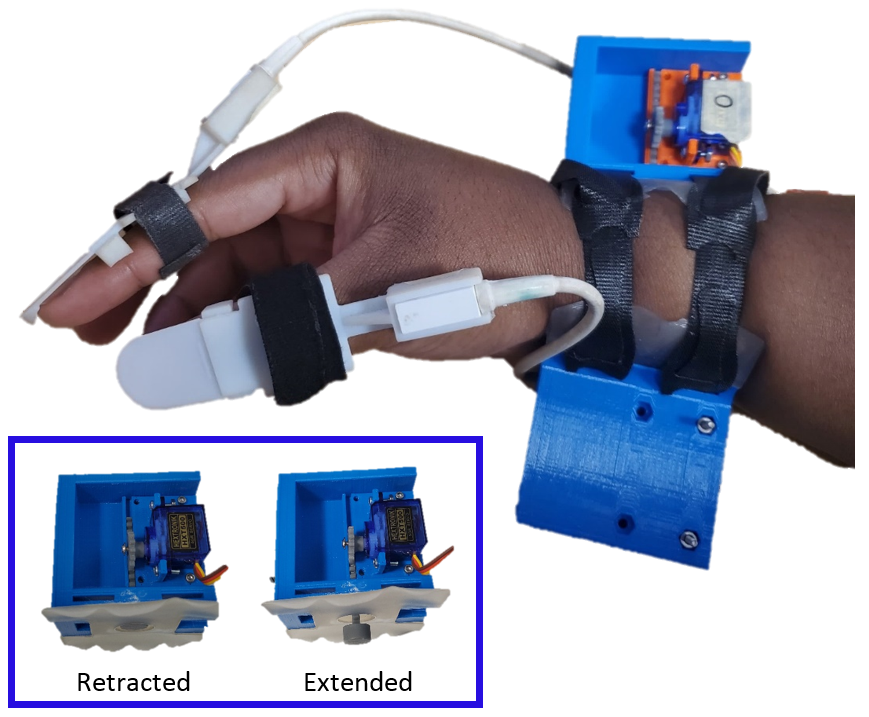}
  \caption{The user wears a pair of 1-DoF wrist-worn tactile haptic devices on the ventral and the dorsal side of their wrist and finger tracking sensors mounted on the index finger and thumb. Each wrist-worn device provides skin deformation normal to the user’s wrist via a tactor connected to a rack and pinion and driven by a servo motor.
}
  \label{fig:deviceFig}
\end{figure}

The user wears two \newtext{of these} devices simultaneously on the same wrist, one on the ventral side, and the other one on the dorsal side of the wrist as shown in Fig.~\ref{fig:deviceFig}. The poses of the user's index finger and thumb are tracked using an electromagnetic 6-DoF tracker by Ascension trakSTAR with two Model 800 sensor\newtext{s} and a pair of 3D-printed mounts on the dorsal (fingernail) side of each finger, leaving the \newtext{ventral (palm-facing)} side of the fingertips free during manipulation tasks. The tracked finger pose is displayed in the virtual environment (as will be discussed in Section \ref{vr_environment_sec}) so the user can interact with a dynamic virtual object. Each device provides haptic feedback corresponding to virtual interaction forces occurring on either the index finger and/or the thumb (as will be described in Section \ref{mapping_sec}). 

\subsection{Virtual Environment} 
\label{vr_environment_sec}

\newtext{To connect our physical devices to our virtual world, we apply a Hooke's Law model ($|F|=k|x|$) to convert the} interaction force \newtext{($F = F_x+F_y+F_z$)} in the virtual environment to a desired position of the tactor, which pushes onto the skin. \newtext{The applied force \newtext{(our interaction force)}, \emph{$|F|$}, represents the magnitude of the interaction forces at the fingertips. The effective stiffness, \emph{k}, is a manually tuned scaling factor discussed below. The penetration or displacement, \emph{$|x|$}, represents the tactor displacement.} 

\begin{figure}[t]
\vspace{1.5mm}
  \centering
  \includegraphics[width=\columnwidth]{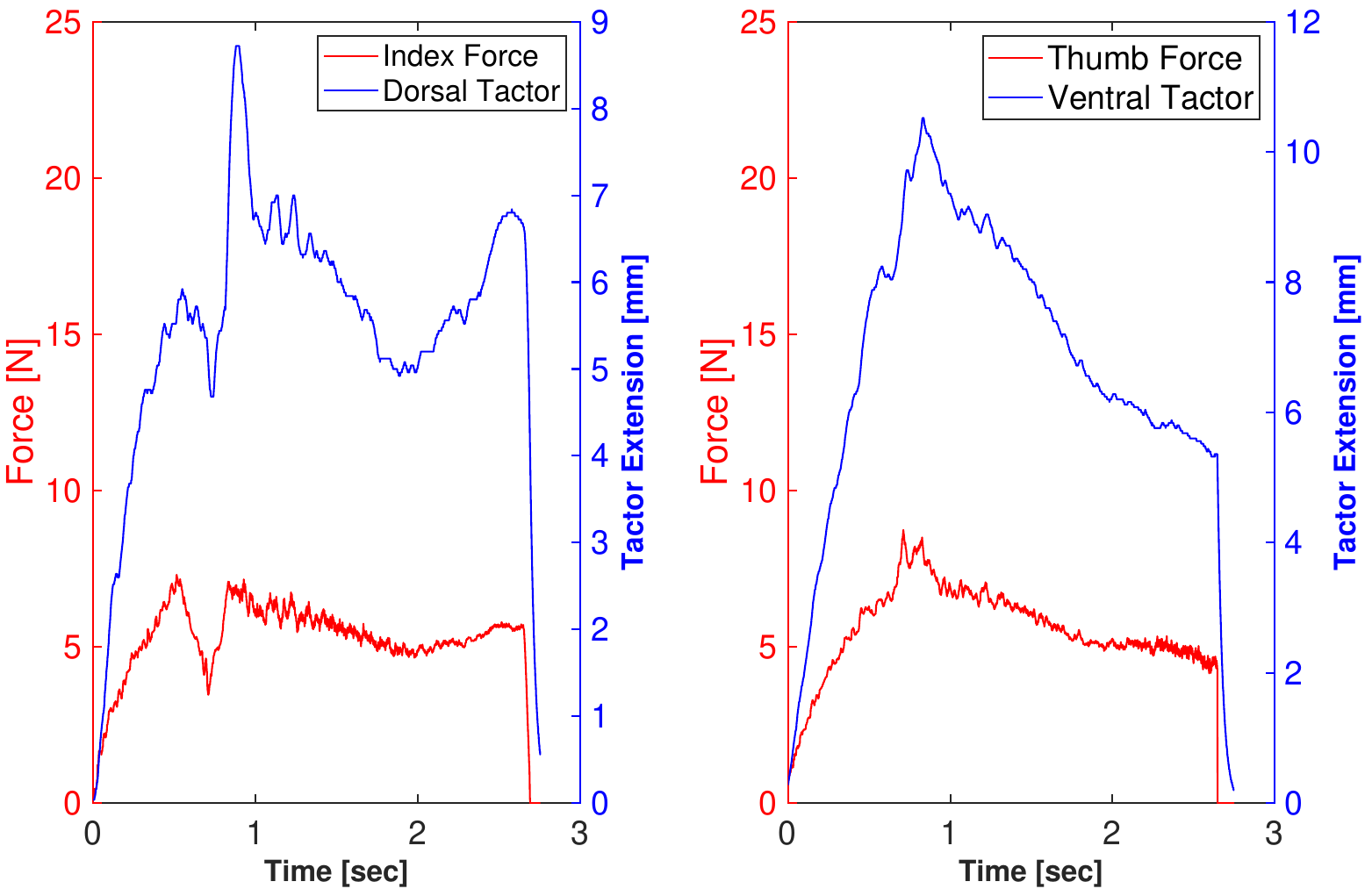}
  \caption{Shown are examples of the (left) virtual index finger force being mapped to a tactor position on the dorsal side of the wrist and (right) virtual thumb force being mapped to a tactor position on the ventral side of the wrist during grasp of a virtual cube. The \newtext{magnitude of the} interaction force \newtext{of each finger} measured in the virtual environment \newtext{is scaled down by a factor of 0.40.}}
  \label{fig:force_to_tactor_pos}
\end{figure}

\newtext{The device’s commanded tactor extension was scaled down by a factor of 0.40 relative to the magnitude of the calculated interaction forces to \newtext{prevent the tactor from being commanded beyond its usable workspace}. We determined this value via informal pilot testing in order to maintain user comfort and to ensure that most virtual interactions resulted in tactor displacement that was within the workspace of the wrist-worn devices. This scaling factor remained constant for all participants during the study.} \newtext{An example of the scaling between the interaction force and tactor extension during a trial is} shown in Fig.~\ref{fig:force_to_tactor_pos}.

\begin{figure}[t]
\vspace{1.5mm}
  \centering
  \includegraphics[width=\columnwidth]{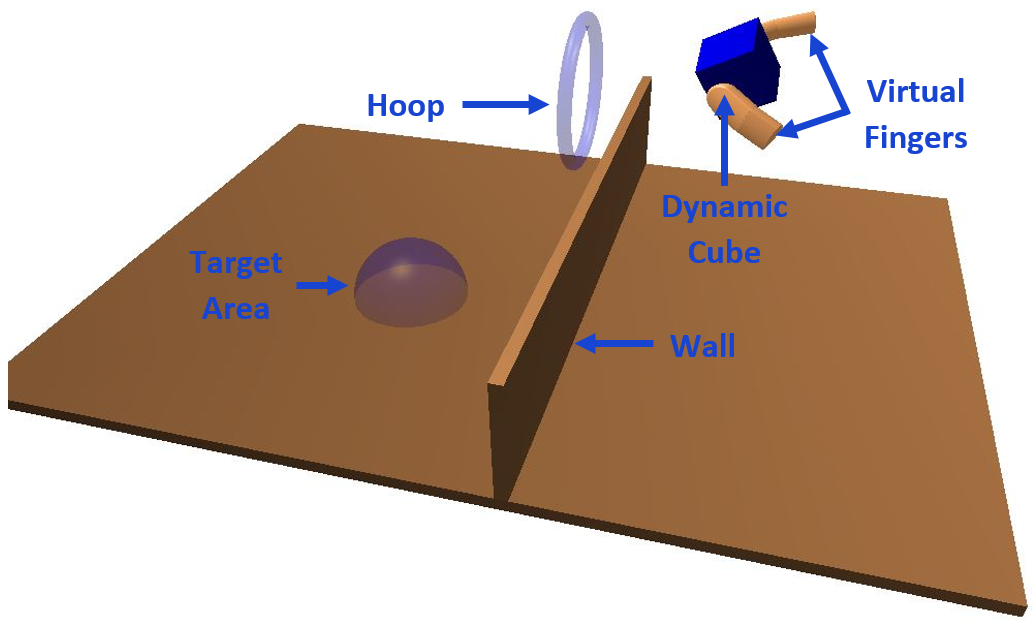}
  \caption{The virtual environment and the \newtext{pick-and-place} task designed for the experiment: the user interacts with the dynamic cube via finger-shaped meshes representing the user's index finger and thumb. After grasping the cube, the user guides it through the hoop and into the target area. The wall is an intentionally placed obstacle to encourage users to lift the cube off the virtual ground.}
  \label{fig:vr_environmentFig}
\end{figure}

The virtual environment and all interaction forces are simulated using the CHAI3D haptics and simulation framework ~\cite{Conti03}. The virtual environment contains a single dynamic cube, a wall, a hoop, and an ellipsoid target area as shown in Fig. \ref{fig:vr_environmentFig}. The wall is an intentionally placed obstacle to encourage users to lift the cube off the virtual ground before guiding it through the hoop and into the target area. The users can interact with the environment via \newtext{virtual avatars in the form of} finger-shaped meshes representing the user's index finger and thumb. Upon contact, the servo motors are given a position command \newtext{equal to the magnitude of} calculated interaction forces \newtext{times the scaling factor of 0.40}. 

\subsection{Finger-to-Tactor Mappings}
\label{mapping_sec}

\begin{figure}[t]
\vspace{1.5mm}
  \centering
  \includegraphics[width=\columnwidth]{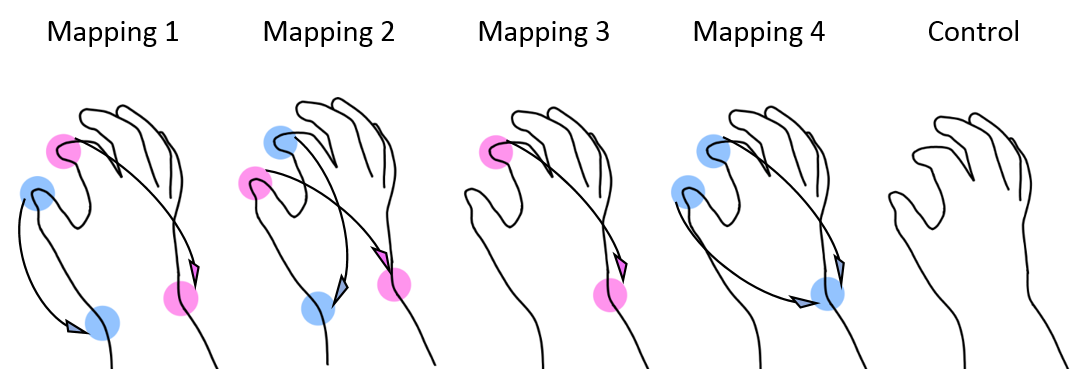}
  \caption{Diagram \newtext{of} the 5 finger-to-tactor mappings: (1) thumb mapped to the ventral side of the wrist and index finger to the dorsal side, (2) thumb mapped to dorsal side of the wrist and index finger to the ventral side, (3) index finger mapped to the dorsal side of the wrist while the thumb is neglected, (4) average of both fingers mapped to the dorsal side of the wrist, and \newtext{control --} no haptic feedback.}
  \label{fig:mappingsFig}
\end{figure}

We considered three main concepts for rendering haptic feedback on the wrist: (1) haptic feedback is rendered on the dorsal and ventral sides of the wrist independently (depending on different finger-to-tactor mappings defined as Mapping 1 and Mapping 2), (2) haptic feedback is rendered only on the dorsal side of the wrist (depending on different finger-to-tactor mappings defined as Mapping 3 and Mapping 4), and (3) no haptic feedback is rendered (\newtext{Control}). Fig. \ref{fig:mappingsFig} shows the five finger-to-tactor mappings used in our study.


For Mapping 1, the index finger force is mapped to the dorsal side of the wrist and the thumb force is mapped to the ventral side of the wrist. This seemed logical because of geometric congruence, which might make it easy for users to interpret wrist haptic feedback corresponding to finger haptic feedback. For Mapping 2, we switched the wrist locations compared to Mapping 1. This was motivated \newtext{by biomechanical considerations and expanding on our previous work~\cite{Sarac2022}.} Index finger movement corresponds to \newtext{musculature} on the ventral side of the \newtext{forearm such as the flexor digitorum profundus} and thumb movement corresponds to \newtext{musculature on} the dorsal side of the\newtext{ forearm such as the abductor pollicis longus. It stands to reason that} pressing on or near \newtext{these structures} could also generate a mapping that is natural for users to understand. \newtext{Additionally, there may be an effect of muscle activity on the wrist, where tactors have been placed in previous studies~\cite{Sarac2022}.}  

Mappings 3 and are 4 are motivated by the idea that it may be possible to reduce \newtext{actuation} complexity by using only a single tactor. \newtext{For} Mapping 3, only the \newtext{interaction forces acting on the} index finger \newtext{are considered}, and it is mapped to the dorsal side of the wrist. \newtext{For} Mapping 4, we averaged the forces of the two fingers and mapped that average to the dorsal side of the wrist. For both Mappings 3 and 4, we could have mapped to the ventral side of the wrist, \newtext{but} prior work~\cite{sarac2019haptic} indicated that either side would be effective. \newtext{Finally, ``Control" had no programmed} haptic feedback, which was important as a baseline to measure the effect of \newtext{the other mappings}. \newtext{For all mappings, the mounting of the device on the user's wrist remained consistent regardless of the actuator activity.}

\section{Experimental Methods} \label{userStudySec}

\subsection{Procedures}

To evaluate the finger-to-tactor mappings, we conducted a user study based on a \newtext{simple two-finger} pick-and-place task. \newtext{Participants were instructed} to grasp the cube \newtext{with only their index finger and thumb}, lift it, \newtext{guide the cube} through the hoop (only the center of the cube is required to pass through the hoop), and place the cube in the target area (only the center of the cube is required to be placed in the target area).

20 participants (ages 19 to 35, 7 females, 13 males, 19 right-handed, 1 left-handed) joined the study. The Stanford University Institutional Review Board approved the experimental protocol and all participants gave informed consent. All participants performed the manipulation tasks using their dominant hand to eliminate the impact of handedness. Participants were split into 2 groups of equal size \newtext{which will be referred to as the \emph{visible} cube group and the \emph{invisible} cube group.}

The \newtext{visible cube} group performed all tasks with the virtual cube being visible to them \newtext{throughout the experiment}. 

The \newtext{invisible cube} group performed the tasks with the added condition that when they picked up the cube \newtext{by 0.01~m vertically with respect to the virtual ground}, it became invisible to them. \newtext{However, the user continued to receive haptic feedback based on their dynamic interactions. For instance, if the user opened their fingers such that the cube fell to the virtual ground, they received decreased and ultimately zero haptic feedback accordingly. The visual feedback of the cube was determined by the vertical position of the cube and therefore remained invisible initially when the cube began to fall but became visible as the cube’s vertical position became within 0.01 m of the virtual ground. For the invisible cube group, their virtual finger avatars remained visible throughout the experiment.} 

\begin{figure}[t]
  \vspace{1.5mm}
  \centering
  \includegraphics[width=\columnwidth]{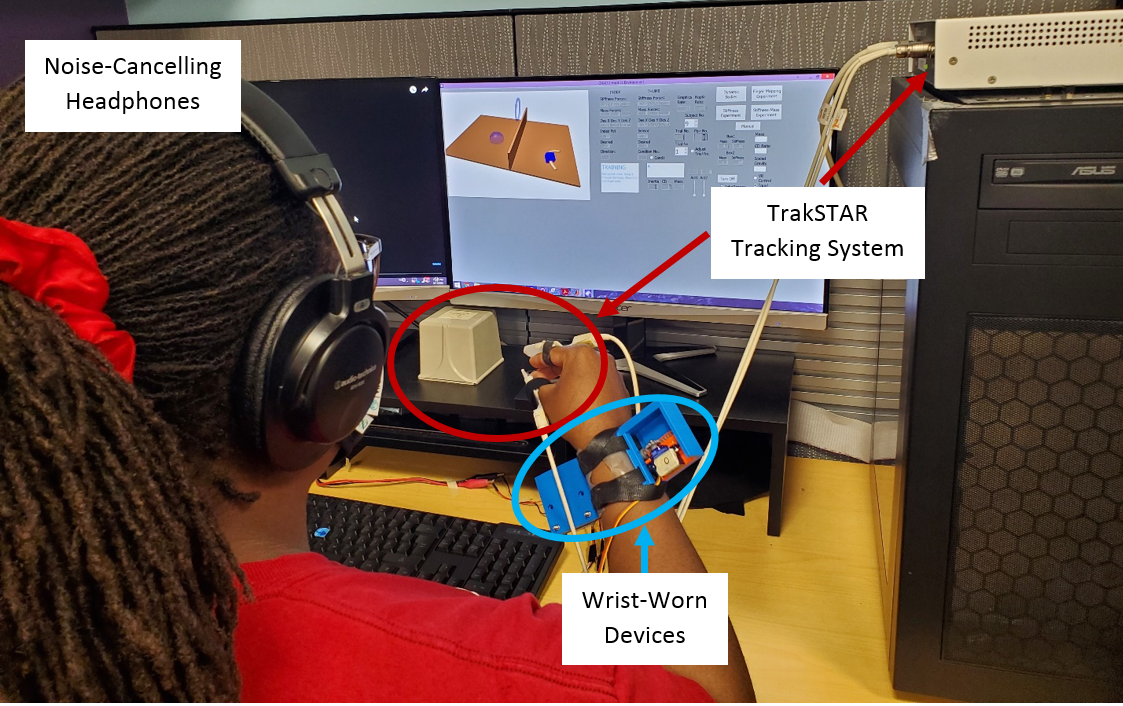}
  \caption{Experimental task: a user performs the pick-and-place task while receiving haptic feedback via  \newtext{a pair of} wrist-worn devices. \newtext{The user's} fingers are tracked by the Ascension trakSTAR electromagnetic tracking system's 3 main components: the finger-mounted sensors placed on the user's index finger and thumb, the mid-range transmitter located near the monitor, and the desktop electronics unit with integrated power supply on the right.}
  \label{fig:experimentSetupFig}
\end{figure}

Haptic feedback conditions and all material properties of the virtual objects were \newtext{otherwise} identical for both groups. Participants wore \newtext{finger-mounted sensors placed on the index finger and thumb to track the pose of the virtual finger avatars. Participants also wore} headphones playing white noise throughout the experiment to \newtext{avoid being biased by auditory cues by} \newtext{canceling the noise from the haptic devices' servo motors and transmissions as shown in Fig. \ref{fig:experimentSetupFig}}. 

\begin{figure}[t]
  \centering
  \includegraphics[width=\columnwidth]{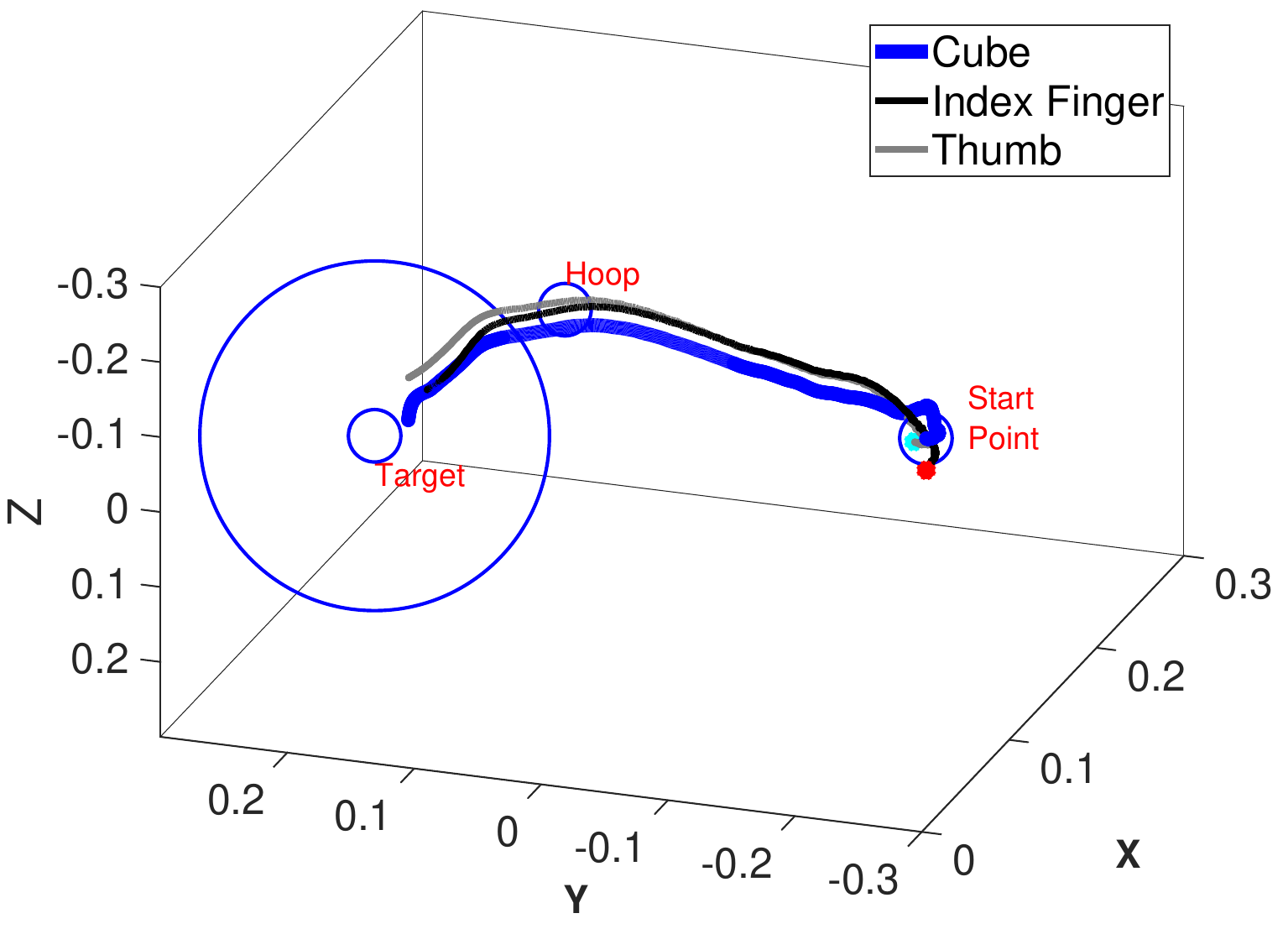}
  \caption{A representative path taken by the user's index finger, thumb, and the cube during the pick-and-place task. This \newtext{kinematic} data \newtext{was} collected during the user's attempt of the pick-and-place task in the testing phase with \newtext{the} Mapping 1 haptic condition.}
  \label{fig:kinematicDataFig}
\end{figure}

The experiment had three phases: pre-trial, training, and testing. Participants were instructed to \newtext{perform the pick-and-place task} as quickly and accurately as possible without dropping the cube \newtext{as shown in Fig. \ref{fig:kinematicDataFig}}. When participants touched the cube, they received different types of haptic feedback \newtext{via} normal indentation into the skin on the wrist, or possibly no feedback at all, to assist with the virtual manipulation. Participants were given a visual cue \newtext{that} they guided and placed the cube properly \newtext{through the hoop and then the target area} by changing the \newtext{opacity} from translucent to opaque. \newtext{If the user incorrectly skipped the hoop, neither the hoop nor target area would become opaque and the experiment would not proceed until the cube completed its proper path.} 

During the pre-trial phase, participants were given an opportunity to briefly explore the virtual environment freely \newtext{without haptic feedback}. The training phase consisted of 5 trials, which provided an example of each of the 5 mappings. All participants completed the training phase with mappings presented in the same order: Mapping 1, Mapping 2, Mapping 3, Mapping 4, and Control (with no haptic feedback) (see Fig. \ref{fig:mappingsFig}). During the testing phase, each participant was assigned a pseudo-random order \newtext{based on Latin Squares in the reduced form} in which mappings were presented to them. They performed the pick-and-place task \newtext{in blocks of 10 trials with the same mapping, with minimum 1-minute breaks in between trial blocks to avoid fatigue.} \newtext{This was repeated until the participant completed the tasks for all mappings, for a} total of 50 trials \newtext{per participant}.

\subsection{Metrics}

We evaluated user performance using three metrics: task completion time, path length, and interaction forces at the fingertips. The task completion time is defined as the \newtext{elapsed} time needed to complete the task successfully \newtext{beginning from the instant the user makes initial simultaneous contact of the index finger and thumb with the virtual cube until the instant the virtual cube is correctly placed in the target area}. The path lengths are the summation of the displacements of the index finger, thumb, and virtual cube throughout a trial. Finally, interaction forces indicate the local forces rendered at the fingertips in the normal and shear directions to the virtual fingertips. Participants were also given a survey to rate the difficulty of completing the task with the different mappings.

\section{Results and Discussion}

\subsection{Learning}
 
We examined the potential effects of learning in terms of path length and task completion time, regardless of the mapping. When looking at participants' progression through the experiment (i.e. comparing their performance at the first and the last trials of the experiment), we observe that there is not a noticeable learning curve in moving the virtual fingers and cube along the most efficient path \newtext{in terms of path length and completion time} to complete the pick-and-place task (data not shown). We conclude that \newtext{because} the task was simple and the visual and/or haptic feedback sufficiently clear, that any learning likely occurred \newtext{during} the training \newtext{phase of the experiment}.


\subsection{Task Completion Time and Path Length}
\label{comptetion_time_path_length_sec}

\begin{figure}[t]
  \centering
  \includegraphics[width=\columnwidth]{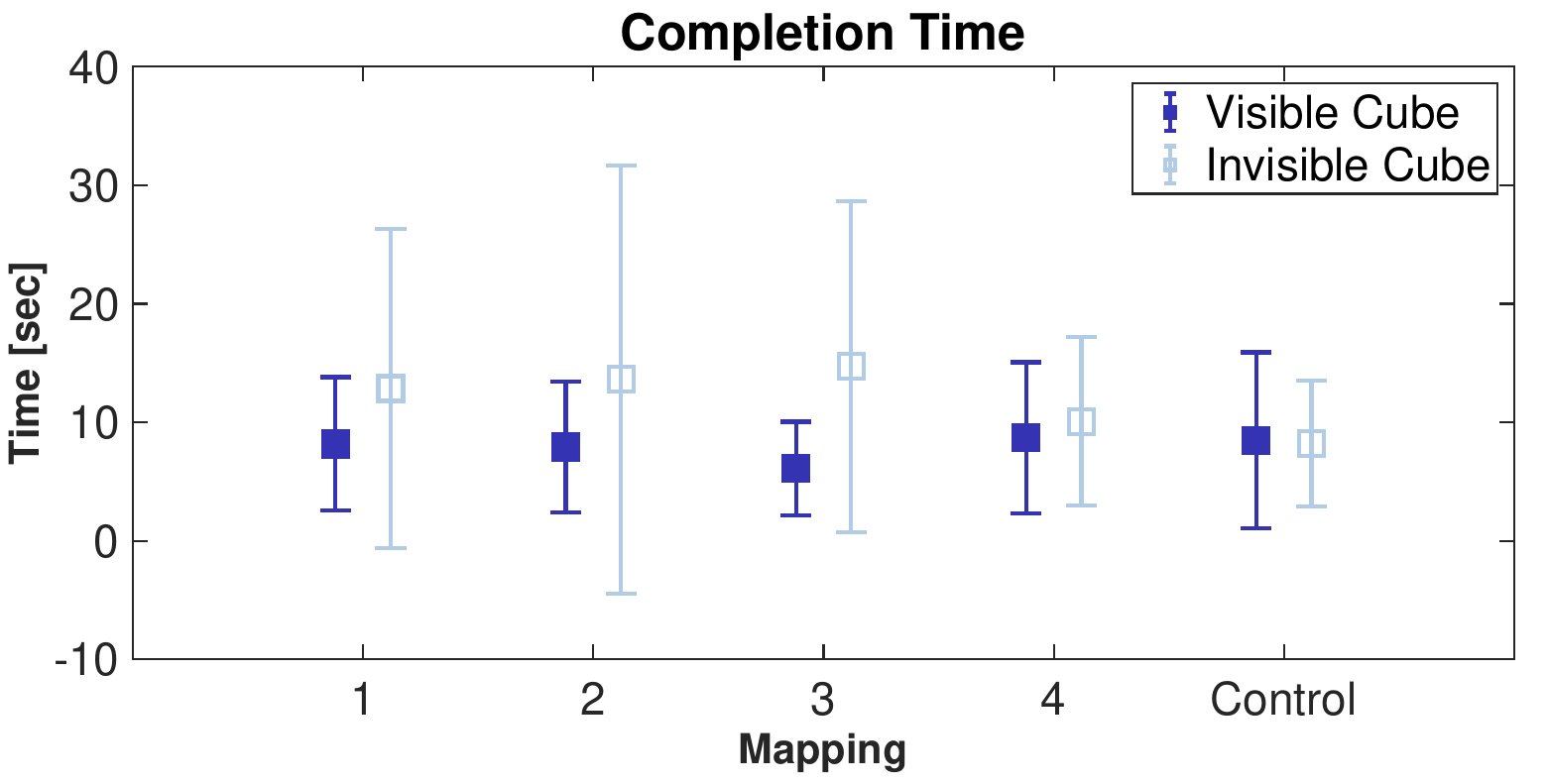}
  \caption{Mean completion time \newtext{with standard deviation} of the pick-and-place task over 10 trials for 5 finger-to-tactor mappings.}
  \label{completionTimeFig}
\end{figure}

\begin{figure}[t]
  \vspace{1.5mm}
  \centering
  \includegraphics[width=\columnwidth]{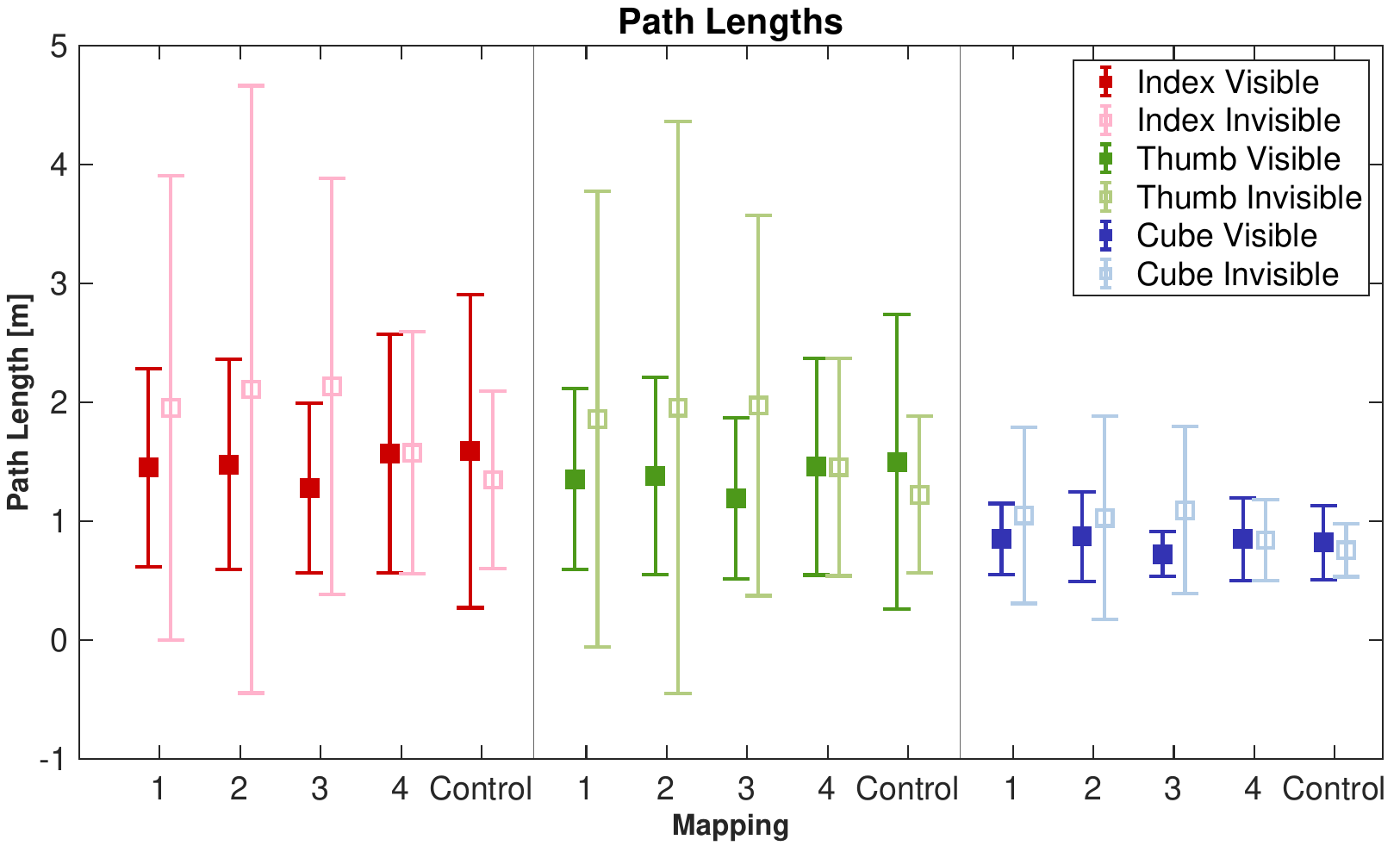}
  \caption{Mean path lengths \newtext{with standard deviation} for the index finger, the thumb, and the cube during the pick-and-place task over 10 trials for 5 finger-to-tactor mappings.}
  \label{pathLengthFig}
\end{figure}

\newtext{We performed \newtext{2-way ANOVA tests with independent variables of haptic mapping (5 levels) and visibility (2 levels)} and calculated p-values and interaction effects at 95\% confidence as shown in Table \ref{resultsTable}, with statistically significant items highlighted in yellow. We observed no statistically significant difference in task completion time or path length regardless of mapping for all subjects. However, we observed a significant difference between the visible cube group and invisible cube group.}  When the cube is visible, subject took less time to complete the task and used shorter paths to complete the task as shown in Figs.\ \ref{completionTimeFig} and \ref{pathLengthFig}. This shows that participants found the task easier with the visible cube. 


\begin{table}[h]
\caption{\newtext{ANOVA Results for Independent Variables of\\ Mapping and Visibility}}
\label{resultsTable}
\centering
\resizebox{\linewidth}{!}{%
\begin{tabular}{|c|c|c|c|} 
\hline
~                                                                             & \begin{tabular}[c]{@{}c@{}}\textbf{p-value by}\\\textbf{Mapping}\end{tabular} & \begin{tabular}[c]{@{}c@{}}\textbf{p-value by}\\\textbf{Visibility}\end{tabular} & \begin{tabular}[c]{@{}c@{}}\textbf{Interaction}\\\textbf{Effects}\end{tabular}  \\ 
\hline
\begin{tabular}[c]{@{}c@{}}\textbf{Completion} \\\textbf{Time}\end{tabular}   & 0.6726                                                                        & {\cellcolor{yellow}}0.0007                                                                           & 0.1309                                                                          \\ 
\hline
\begin{tabular}[c]{@{}c@{}}\textbf{Index Path} \\\textbf{Length}\end{tabular} & 0.7411                                                                        & {\cellcolor{yellow}}0.0318                                                                           & 0.1867                                                                          \\ 
\hline
\begin{tabular}[c]{@{}c@{}}\textbf{Thumb Path}\\\textbf{Length}\end{tabular}  & 0.7312                                                                        & {\cellcolor{yellow}}0.0435                                                                           & 0.1764                                                                          \\ 
\hline
\begin{tabular}[c]{@{}c@{}}\textbf{Cube Path } \\\textbf{Length}\end{tabular} & 0.3268                                                                        & {\cellcolor{yellow}}0.02623                                                                          & 0.1406                                                                          \\ 
\hline
\begin{tabular}[c]{@{}c@{}}\textbf{Index Shear} \\\textbf{Force}\end{tabular} & 0.5366                                                                        & {\cellcolor{yellow}}0.0007                                                                           & 0.7636                                                                          \\ 
\hline
\begin{tabular}[c]{@{}c@{}}\textbf{Index Normal}\\\textbf{Force}\end{tabular} & 0.1724                                                                        & {\cellcolor{yellow}}0.0192                                                                           & 0.3534                                                                          \\ 
\hline
\begin{tabular}[c]{@{}c@{}}\textbf{Thumb Shear}\\\textbf{Force}\end{tabular} & 0.3994                                                                        & {\cellcolor{yellow}}$<$0.0001                                                                           & 0.2672                                                                          \\ 
\hline
\begin{tabular}[c]{@{}c@{}}\textbf{Thumb Normal}\\\textbf{Force}\end{tabular}  & {\cellcolor{yellow}}0.0151                                                    & 0.0504                                                                                               & 0.0989                                                                          \\
\hline
\end{tabular}
}
\end{table}

\subsection{Interaction Forces}
 
\newtext{We evaluated the normal and shear components of the interaction forces in order to better explain the behavior of the manipulation task. The normal component is related to the stiffness of the cube and how much the cube is squeezed statically. Once the cube is squeezed, the normal component remains essentially constant regardless the motions of the cube. The shear component is a function of (1) the surface friction and how much the finger slides across the surface and (2) the cube's mass and the user's acceleration. Due to its dynamics nature, its magnitude is much higher than the normal component~\cite{lederman1993extracting}.} 
 
\begin{figure}[t]
  \vspace{1.5mm}
  \centering
  \includegraphics[width=\columnwidth]{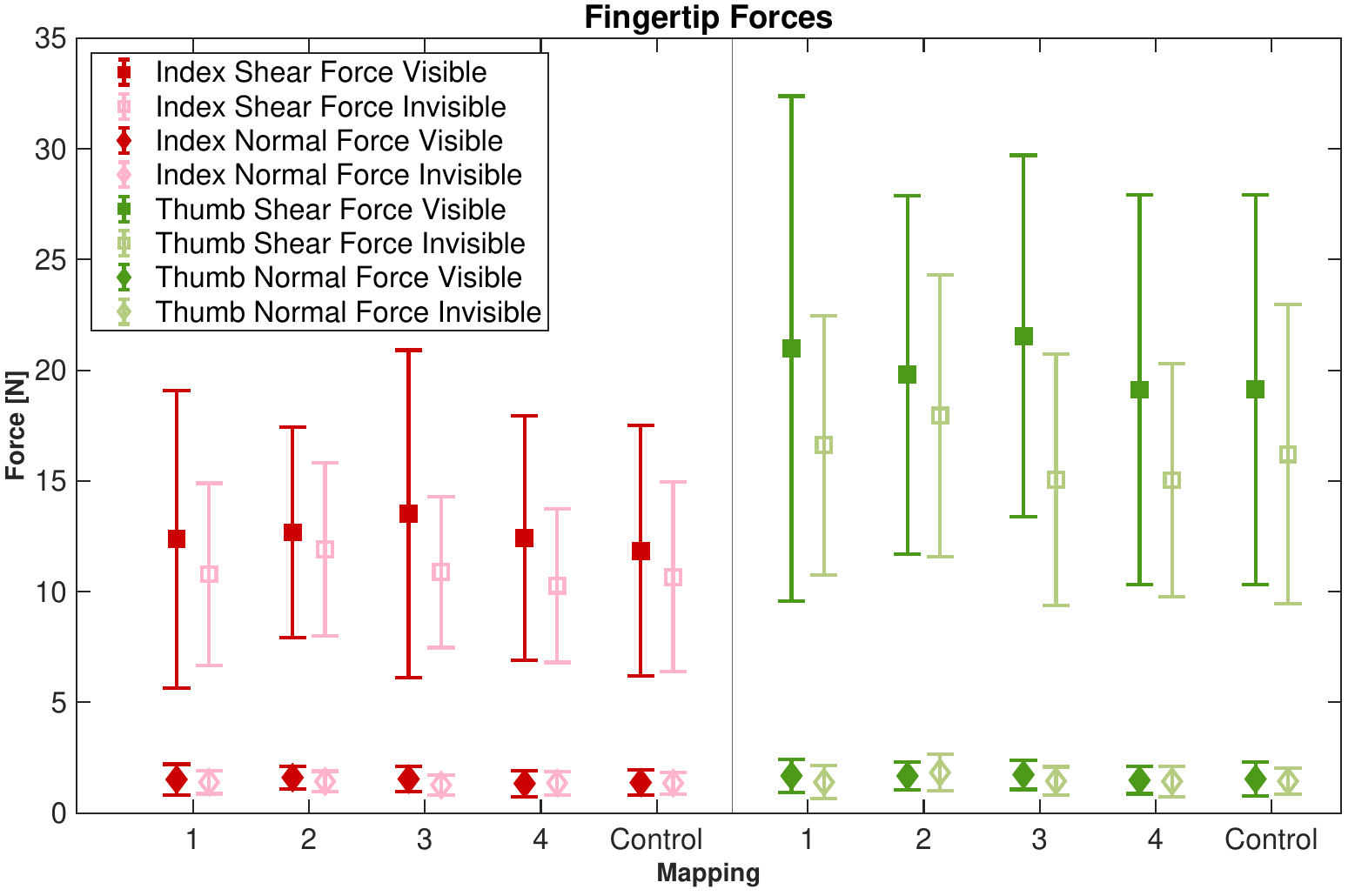}
  \caption{Mean finger-to-cube interaction forces \newtext{with standard deviation} in the normal and the shear directions separately during the pick-and-place task over 10 trials for 5 finger-to-tactor mappings.}
  \label{normalAndShearForcesFig}
\end{figure}

\newtext{We performed the same \newtext{2-way} ANOVA tests with independent variables of haptic mapping and visibility, with 5 and 2 levels, respectively, and presented the results in Table \ref{resultsTable}. We observed no statistically significant difference in the normal and shear components of the index forces and in the shear component of the thumb forces, regardless of mapping for all subjects. We did observe statistical significance between mapping conditions in the normal component of the thumb forces ($p = 0.0151$). Additionally, we observed a significant difference between the visible cube group and invisible cube group for all fingers and force components except for the normal component of the thumb forces ($p = 0.0504$).} For these magnitudes of interaction forces at the fingertip, we observe that forces were larger for the visible cube group \newtext{as shown in Fig. \ref{normalAndShearForcesFig}}, perhaps because participants are more confident/less cautious and therefore applied more shear force to the cube. 

The difference between shear forces with the visible and invisible cube is smallest \newtext{for Mapping 2 and largest for Mapping 3 as shown in Fig. \ref{normalAndShearForcesFig}}. \newtext{For the index finger, the difference in mean shear forces was  1.5789 N, 0.7640 N, 2.6365 N, 2.1436 N, 1.1840 N, for Mappings 1, 2, 3, 4, and Control, respectively. For the thumb, the difference in mean shear forces was 4.3668 N, 1.8460 N, 6.4848 N, 4.0846 N, 2.9248 N, for Mappings 1, 2, 3, 4, and Control, respectively.}

\subsection{Survey Results}
Participants were asked to rate the difficulty of the pick-and-place task for each mapping as very easy, easy, moderate, hard, or very hard. To determine which mappings were perceived as most and least difficult, we converted the ratings to values ranging from 1 to 5 corresponding to very easy and very hard, respectively. The number ratings for each mapping were averaged to compare overall ratings between mappings. The results indicate that Mapping 1 allowed participants to perform the task with most ease and Mapping 4 with the most difficulty. Mappings 2, 3, and the Control had similar overall ratings, slightly below that of Mapping 4. \newtext{We performed a non-parametric Friedman test on these ratings and observed no statistically significant preference of one mapping over any other.}

Participants were also asked to qualitatively describe what strategies they used to perform the task with different mappings. Most participants relied on visual cues and feedback when haptic feedback was not available and stated that haptic feedback was generally helpful in performing the pick-and-place task for most mappings. Although some participants found the tactile sensation from the device to be unnatural, most indicated that the haptic feedback was especially useful in quickly knowing when the virtual object was picked up and dropped. Some participants also relied on other methods to secure the virtual object between the virtual thumb and index finger such as moving slowly and steadily and thinking about the position of the fingertips relative to each other.

\section{Conclusions and Future Work}

In summary, we found that the order in which users were presented with each of the 5 finger-to-tactor mappings does not result in a noticeable difference in the task performance metrics. However, when changing the visibility of the cube, the task performance metrics had noticeable trends. Participants also \newtext{qualitatively} rated \newtext{Mapping 1, the geometrical congruent mapping,} as the best for ease of task performance \newtext{although this preference was not statistically significant}.

Completion time for each trial and path lengths of the index finger, thumb, and cube were lower for the \newtext{visible cube group} compared to the \newtext{invisible cube group} during the experiment, showing more efficient completion of the task \newtext{for the former group. However, no statistically significant differences in performance occurred regardless of mapping.}

\newtext{We observed no statistically significant difference in index normal, index shear, and thumb shear force regardless of mapping for all subjects. However, we did observe statistical significance between mappings for thumb normal force. Additionally, we observed a significant difference between the visible cube group and invisible cube group when analyzing the effect of visibility for the index normal, index shear, and thumb shear force. However, there was \newtext{no statistically significant difference} for the thumb normal force. All performance metrics were evaluated during the testing phase of the experiment.} 
 
In future work, we plan to develop and test a multi-DoF device and determine whether adding more DoFs to the haptic feedback will be as effective on the wrist as it is on the fingertips, as well as which combination of DoFs are optimal. We believe that there will be an improvement over 1-DoF haptic feedback on the wrist, and will quantify the differences in performance metrics and behavior in the virtual environment. We also intend to extend this work to \newtext{a more immersive} mixed reality \newtext{environment} and continue to use the multi-DoF device and mappings we determine are optimal. A new mixed reality environment will be developed using the Unity game engine and CHAI3D haptics and simulation framework. We will integrate this new framework with a Microsoft HoloLens 2 Augmented Reality headset. In a future user study, we will investigate how the augmented haptic feedback aids in training and perception for virtual objects integrated in a real-world environment.

\addtolength{\textheight}{-12cm}   




\bibliographystyle{IEEEtran}
\bibliography{myReferences}

\begin{thebibliography}{10}
\providecommand{\url}[1]{#1}
\csname url@rmstyle\endcsname
\providecommand{\newblock}{\relax}
\providecommand{\bibinfo}[2]{#2}
\providecommand\BIBentrySTDinterwordspacing{\spaceskip=0pt\relax}
\providecommand\BIBentryALTinterwordstretchfactor{4}
\providecommand\BIBentryALTinterwordspacing{\spaceskip=\fontdimen2\font plus
\BIBentryALTinterwordstretchfactor\fontdimen3\font minus
  \fontdimen4\font\relax}
\providecommand\BIBforeignlanguage[2]{{%
\expandafter\ifx\csname l@#1\endcsname\relax
\typeout{** WARNING: IEEEtran.bst: No hyphenation pattern has been}%
\typeout{** loaded for the language `#1'. Using the pattern for}%
\typeout{** the default language instead.}%
\else
\language=\csname l@#1\endcsname
\fi
#2}}

\bibitem{schorr2017fingertip}
S.~B. Schorr and A.~M. Okamura, ``Fingertip tactile devices for virtual object
  manipulation and exploration,'' in \emph{Conference on Human Factors in
  Computing Systems}, 2017, pp. 3115--3119.

\bibitem{Leonardis2017}
D.~{Leonardis}, M.~{Solazzi}, I.~{Bortone}, and A.~{Frisoli}, ``A {3-RSR}
  haptic wearable device for rendering fingertip contact forces,'' \emph{IEEE
  Transactions on Haptics}, vol.~10, no.~3, pp. 305--316, 2017.

\bibitem{Chinello2018}
F.~Chinello, C.~Pacchierotti, M.~Malvezzi, and D.~Prattichizzo, ``A novel
  {3-RRS} wearable fingertip cutaneous device for virtual interaction,'' in
  \emph{Haptic Interaction}, 2018, pp. 63--65.

\bibitem{SalvatoRAL2022}
M.~Salvato, N.~Heravi, A.~M. Okamura, and J.~Bohg, ``Predicting hand-object
  interaction for improved haptic feedback in mixed reality,'' \emph{IEEE
  Robotics and Automation Letters}, vol.~7, no.~2, pp. 3851--3857, 2022.

\bibitem{Pezent2019}
E.~Pezent, A.~Israr, M.~Samad, S.~Robinson, P.~Agarwal, H.~Benko, and
  N.~Colonnese, ``Tasbi: {M}ultisensory squeeze and vibrotactile wrist haptics
  for augmented and virtual reality,'' in \emph{IEEE World Haptics Conference},
  2019, pp. 1--6.

\bibitem{Young2019}
E.~M. Young, A.~H. Memar, P.~Agarwal, and N.~Colonnese, ``Bellowband: {A}
  pneumatic wristband for delivering local pressure and vibration,'' in
  \emph{IEEE World Haptics Conference}, 2019, pp. 55--60.

\bibitem{Moriyama2018}
T.~K. {Moriyama}, A.~{Nishi}, R.~{Sakuragi}, T.~{Nakamura}, and H.~{Kajimoto},
  ``Development of a wearable haptic device that presents haptics sensation of
  the finger pad to the forearm,'' in \emph{IEEE Haptics Symposium}, 2018, pp.
  180--185.

\bibitem{RaitorICRA2017}
M.~Raitor, J.~M. Walker, A.~M. Okamura, and H.~Culbertson, ``{WRAP: W}earable,
  restricted-aperture pneumatics for haptic guidance,'' in \emph{IEEE
  International Conference on Robotics and Automation}, 2017, pp. 427--432.

\bibitem{Johansson2009}
R.~S. Johansson and J.~R. Flanagan, ``Coding and use of tactile signals from
  the fingertips in object manipulation tasks,'' \emph{Nature Reviews
  Neuroscience}, vol.~10, no.~5, pp. 345--359, 2009.

\bibitem{CulbertsonAR2018}
H.~Culbertson, S.~Schorr, and A.~M. Okamura, ``Haptics: The present and future
  of artificial touch sensations,'' \emph{Annual Review of Control, Robotics,
  and Autonomous Systems}, vol.~1, pp. 385--409, 2018.

\bibitem{sarac2019effects}
M.~Sarac, A.~M. Okamura, and M.~Di~Luca, ``Effects of haptic feedback on the
  wrist during virtual manipulation,'' \emph{arXiv preprint arXiv:1911.02104},
  2019.

\bibitem{sarac2019haptic}
------, ``Haptic sketches on the arm for manipulation in virtual reality,''
  \emph{arXiv preprint arXiv:1911.08528}, 2019.

\bibitem{Sarac2022}
M.~Sarac, T.~M. Huh, H.~Choi, M.~Cutkosky, M.~D. Luca, and A.~M. Okamura,
  ``Perceived intensities of normal and shear skin stimuli using a wearable
  haptic bracelet,'' \emph{IEEE Robotics and Automation Letters}, 2022, in
  press, DOI: 10.1109/LRA.2021.3140132.

\bibitem{Conti03}
F.~Conti, F.~Barbagli, R.~Balaniuk, M.~Halg, C.~Lu, D.~Morris, L.~Sentis,
  J.~Warren, O.~Khatib, and K.~Salisbury, ``The chai libraries,'' in
  \emph{Proceedings of Eurohaptics 2003}, Dublin, Ireland, 2003, pp. 496--500.

\bibitem{lederman1993extracting}
S.~J. Lederman and R.~L. Klatzky, ``Extracting object properties through haptic
  exploration,'' \emph{Acta Psychologica}, vol.~84, no.~1, pp. 29--40, 1993.

\end{thebibliography}

\end{document}